\documentclass{aastex}

\usepackage{spr-astr-addons}    
\usepackage{txfonts}
\usepackage{graphicx}
\usepackage{epsfig}
\usepackage{natbib}
\usepackage{amssymb}
\usepackage{color}
\usepackage{multirow}
\usepackage[english]{babel}
\def\be{\begin{equation}}
\def\ee{\end{equation}}

\begin{document}

\title{The  variation  of the  fine  structure  constant: testing  the
  dipole model with thermonuclear supernovae}

\shorttitle{Testing a spatial variation of $\alpha$ with SNIa}
\shortauthors{Kraiselburd et al.}

\author{L. Kraiselburd}
\altaffiltext{}{Grupo de Astrof\'{\i}sica, Relatividad y Cosmolog\'{\i}a, 
                Facultad de Ciencias Astron\'{o}micas y Geof\'{\i}sicas, 
                Universidad Nacional de La Plata, 
                Paseo del Bosque S/N 1900 La Plata, 
                Argentina}
\and
\author{S. J. Landau}
\altaffiltext{}{Departamento de F\'{\i}sica, 
                Facultad de Ciencias Exactas y Naturales, 
                Universidad de Buenos Aires}
\altaffiltext{}{IFIBA, CONICET, 
                Ciudad Universitaria - Pab. I, 
                Buenos Aires 1428, 
                Argentina}
\altaffiltext{}{Member of  the Carrera del Investigador Cient\'{\i}fico 
                y Tecnol\'ogico, CONICET}
\email{slandau@df.uba.ar} 
\and 
\author{C. Negrelli}
\affil{Grupo de Astrof\'{\i}sica, Relatividad y Cosmolog\'{\i}a, 
       Facultad de Ciencias Astron\'{o}micas y Geof\'{\i}sicas, 
       Universidad Nacional de La Plata, 
       Paseo del Bosque S/N 1900 La Plata, 
       Argentina}
\and
\author{E. Garc\'\i a--Berro}
\altaffiltext{}{Departament de F\'\i sica Aplicada,
                Universitat Polit\`ecnica de Catalunya,
                c/Esteve Terrades 5,
                08860 Castelldefels,
                Spain}
 \altaffiltext{}{Institute for Space Studies of Catalonia,
                 c/Gran Capita 2--4,
                 Edif. Nexus 201,
                 08034 Barcelona,
                 Spain}

\abstract{The  large-number  hypothesis conjectures  that  fundamental
  constants  may  vary.   Accordingly,  the  space-time  variation  of
  fundamental constants  has been  an active  subject of  research for
  decades.   Recently, using  data  obtained with  large telescopes  a
  phenomenological model  in which  the fine structure  constant might
  vary spatially has been proposed.  We test whether this hypothetical
  spatial  variation  of $\alpha$,  which  follows  a dipole  law,  is
  compatible  with  the  data  of  distant  thermonuclear  supernovae.
  Unlike previous works, in our  calculations we consider not only the
  variation  of the  luminosity distance  when a  varying $\alpha$  is
  adopted, but  we also take  into account  the variation of  the peak
  luminosity  of Type  Ia  supernovae resulting  from  a variation  of
  $\alpha$.  This  is done  using an empirical  relation for  the peak
  bolometric  magnitude  of  thermonuclear supernovae  that  correctly
  reproduces the  results of detailed numerical  simulations.  We find
  that  there  is  no   significant  difference  between  the  several
  phenomenological models studied here and  the standard one, in which
  $\alpha$ does not vary spatially.   We conclude that the present set
  of  data of  Type  Ia  supernovae is  not  able  to distinguish  the
  standard model  from the dipole models,  and thus cannot be  used to
  discard nor to confirm the proposed spatial variation of $\alpha$.}

\keywords{quasars:  absorption lines  --  cosmology: miscellaneous  --
  stars: white dwarfs -- supernovae: general}

\section{Introduction}
\label{Intro}

Since   the   large   number   hypothesis  was   first   proposed   by
\citet{Dirac37}  the  search  for  a  time  variation  of  fundamental
constants has  motivated numerous theoretical and  experimental works.
To  this regard  it is  important to  realize that  the most  commonly
accepted cosmological theories rely on the assumption that fundamental
constants -- like  the gravitational constant $G$,  the fine structure
constant $\alpha$, or the proton-to-electron mass ratio $\mu$\ldots --
are indeed truly and genuinely constant.  However, the assumption that
these  constants  do  not  vary  with  time  or  location  is  just  a
hypothesis, though quite a reasonable an important one, which needs to
be   observationally    corroborated.    Actually,    several   modern
grand-unification  theories predict  that these  constants are  slowly
varying  functions of  low-mass dynamical  scalar fields  -- see,  for
instance,    \cite{Pablo},    \cite{Uzan}   and    \cite{Yuri},    and
re\-fe\-ren\-ces  therein.  In  particular,  the  ongoing attempts  of
unifying  all fundamental  interactions have  led to  the developement
se\-ve\-ral  multidimensional  theories, like  string-motivated  field
theories,  related   brane-world  theories,   and  (related   or  not)
Kaluza-Klein theories, which predict not  only an energy dependence of
the fundamental  constants but also  a dependence of  their low-energy
limits on cosmological times.  Thus, should these theories prove to be
correct,  it is  expected  that fundamental  cons\-tan\-ts would  vary
slowly over  long timescales, or  would vary spatially.  Hence,  it is
natural ask  ourselves which are  the observational consequences  of a
spatio-temporal variation of the  fundamental constants, and to design
new methods  to measure, or  at least to constrain,  such hypothetical
variations, as this  would allow us to confirm or  discard some of the
proposed theories.

According to this theoretical framework,  in the last decade the issue
of the  variation of fundamental  constants has experienced  a renewed
interest, and several observational  stu\-dies have been undertaken to
scrutinize  their  possible  variations  \citep{Uzan,  Yuri},  and  to
establish  constraints on  such variations.   Generally speaking,  the
ex\-pe\-ri\-men\-tal  studies   can  be   grouped  in   two  different
categories, namely  astronomical and  local methods.  The  latter ones
include, among other techniques, geophysical  methods such as the Oklo
natural nuclear reactor that operated about $1.8\times 10^9$ years ago
\citep{Petrov06,Gould06},  the  a\-na\-ly\-sis of  natural  long-lived
$\beta$    decayers   in    geological    minerals   and    meteorites
\citep{Olive04b},  and laboratory  measurements  which compare  clocks
with  different atomic  numbers  \citep{Fischer04,Peik04}. The  former
methods comprise  a large  variety of  methods --  see the  reviews of
\cite{Uzan} and \cite{Yuri} for  extensive dis\-cussions\- of the many
observational  techniques.    However,  the  most   successful  method
employed so  far to  measure hypothetical  variations of  $\alpha$ and
$\mu$ is based on the analysis  of the spectral lines of high-redshift
quasar  absorption   systems,  the  so-called   many-multiplet  method
\citep{Webb99}.  This method compares the characteristics of different
transitions in the same absorption cloud,  and results in a gain of an
order of magnitude in sensibility  respect to previous methods.  As it
should be o\-ther\-wi\-se  expected, most of the  reported results are
consistent with  a null variation of  fundamental constants.  However,
using\- this method \citet{Webb99} and \citet{Murphy03b} have reported
the results of  Keck/HIRES observations which suggest  a smaller value
of  $\alpha$  at high  redshift  as  compared  with its  local  value.
Nevertheless,  an independent  analysis performed  with VLT/UVES  data
gave  null  results  \citep{Srianand04}.   Contrary  to  the  previous
results,  a recent  analysis  using\- VLT/UVES  data  suggests also  a
variation in $\alpha$ but in the oppo\-si\-te sense, that is, $\alpha$
appears to be larger in  the past \citep{Webb11,King12}.  In addition,
it  has been  pointed out  \citep{LS08,KLS13} that  results calculated
from  the mean  value over  a  large redshift  range (or  cosmological
time-scale) are  at variance  with those obtained  considering smaller
intervals.  Thus,  from the  observational point  of view,  a possible
slow variation of fundamental constants with look-back times remains a
controversial  issue,  and  the  discrepancy  between  Keck/HIRES  and
VLT/UVES is yet to be resolved.

Since  the Keck/Hires  and  VLT/UVES observations  rely  on data  from
telescopes  observing  different  hemispheres, it  has  been  recently
suggested that their respective results  can be made consistent if the
fine structure  constant were spatially varying.   Additionally, there
is some recent observational evidence  which could be interpreted as a
hint  for  deviations  from  large-scale  statistical  isotropy.   For
example,  the alignment  of low  multi-poles in  the Cosmic  Microwave
Background  angular  power   spectrum  \citep{anomaliesCMB},  and  the
large-scale   alignment  in   the   QSO   optical  polarization   data
\citep{quasarspol}   may   support   this  explanation.    All   these
observations have  boosted the  interest in the  search for  a spatial
variation of $\alpha$.  As  mentioned, \cite{Webb11} and \cite{King12}
reported a  possible spatial  variation of  $\alpha$, and  showed that
phenomenological  models where  the  variation in  $\alpha$ follows  a
dipole law can  be well fitted to the obtained  data.  This result was
later confirmed  by \cite{Berengut12}.  All these  observational works
also motivated the  theoretical interest in this kind  of studies. For
instance, \citet{mariano} stu\-died if  the reported spatial variation
of $\alpha$  was compatible with  the observations of distant  Type Ia
supernovae (SNIa).  They  did so employing the  Union~2 compilation of
luminosity   distances    \citep{Union2,Suzuki12}.    More   recently,
\citet{YWC14} searched  for a preferred direction  using the Union~2.1
sample and found a preferred  direction which can be well approximated
by a dipole fit.  However, none of these studies took into account the
dependence of the Chandrasekhar limiting  mass on the precise value of
$\alpha$.   The only  study in  which  a dependence  of the  intrinsic
properties  of   Type  Ia  supernovae   has  been  done  is   that  of
\citet{ChibaKohri}.  Specifically,  they a\-na\-ly\-zed the  effect of
changing  $\alpha$  on  the  peak  bolometric  magnitude  of  Type  Ia
supernovae.   However, this  pioneering analysis  only considered  the
dependence of the mean opacity of the expanding photosphere of Type Ia
supernovae on the  value of $\alpha$, and neglected  the dependence of
the Chandraskhar limiting  mass on the precise value  of $\alpha$.  In
this paper  we perform  a similar analysis,  this time  considering as
well the dependence of the  Chandrasekhar mass on $\alpha$.  Thus, our
study complements  and expands that of  \citet{ChibaKohri}. To compare
with observations  we employ the  standard cosmological model  and the
Union~2.1  compilation of  distant  SNIa. Our  paper  is organized  as
follows.  In Sect.~\ref{models}  we explain how our  models are built.
If  follows  Sect.~\ref{resultados},  where we  present  our  results.
Lastly, in Sect.~\ref{discusion} we summarize our main findings and we
present our conclusions.

\section{The luminosity distance relation}
\label{models}

In  this paper,  we use  the measured  luminosity distance  of SNe  Ia
explosions   to   test   the   phenomenological   dipole   models   of
\cite{King12}.   Thermonuclear supernovae  are  best  suited for  this
purpose  as they  are considered  good  standard candles  that can  be
observed up to  very high redshifts.  Actually, SNe  Ia are calibrable
candles, as  its peak luminosity  correlates with the decline  rate of
the  light  curve. This  is  because,  although  the nature  of  their
progenitors  and the  detailed mechanism  of explosion  are still  the
subject of a strong debate,  their observational light curves are well
understood and their individual intrinsic differences can be accounted
for. Hence, observations  of distant SNe Ia are now  used to constrain
cosmological  parameters  \citep{Chiba1,Chiba2},  or  to  discriminate
among  different alternative  cosmological  theories.  However,  their
reliability as distance indicators relies on the assumption that there
is no  mechanism able to  produce an  evolution of the  observed light
curves  over cosmological  distances.   The homogeneity  of the  light
curve  is  essentially due  to  the  homogeneity  of the  nickel  mass
produced  during   the  supernova  outburst  ($M_{\rm   Ni}\sim  0.6\,
M_{\sun}$),  and this  is primarily  determined  by the  value of  the
Chandrasekhar limiting mass, which depends on $\alpha$:
\begin{equation}
M_{\rm Ch}\propto
\left(\frac{e^2}{\alpha G}\right)^{\frac{3}{2}},
\label{chandra}
\end{equation}
where all the symbols have their  usual meaning. Thus, the nickel mass
synthesized    during   the    thermonuclear   outburst    scales   as
$\alpha^{-3/2}$.  Hence, if  $\alpha$ varies so does  the nickel mass,
and  consequently, the  peak  bolometric  luminosity of  thermonuclear
supernovae and  correspondingly the  derived distance. Also,  the peak
luminosity of thermonuclear  supernovae depends on the  opacity of the
expanding  photosphere, that  also  depends on  the  precise value  of
$\alpha$. In the next subsection  we calculate how the peak bolometric
magnitude scales on $\alpha$ taking into account both dependences.

\subsection{The dependence of the peak luminosity on $\alpha$}
\label{sec:mbol}

The  dependence  of the  peak  bolometric  magnitude of  thermonuclear
supernovae  on  $\alpha$  can  be  obtained  using  simple  analytical
arguments. To do  this we follow closely  \cite{ChibaKohri}, this time
taking all the dependencies on  $\alpha$ into account.  To start with,
we recall that the peak luminosity of SNIa is given by:
\begin{equation}
L_{\rm peak}=M_{\rm Ni}q(t_{\rm peak})
\end{equation}
where $M_{\rm Ni}\simeq0.6\,M_{\rm sun}$, and 
\begin{eqnarray}
q(t)=&&\left[S_{\rm Ni}^\beta e^{-t/\tau_{\rm Ni}}+S_{\rm Co}\left(
e^{-t/\tau_{\rm Co}}-e^{-t/\tau_{\rm Ni}}\right)\right]f_{\rm dep}^\gamma(t)+\nonumber\\
&&S_{\rm Co}^\beta\left(e^{-t/\tau_{\rm Co}}-e^{-t/\tau_{\rm Ni}}\right)
\end{eqnarray}
is         the         energy         deposited         by         the
$^{56}$Ni$\rightarrow^{56}$Co$\rightarrow^{56}$Fe  decay chain  inside
the  photosphere  of the  exploding  supernova,  $\tau_{\rm Ni}$,  and
$\tau_{\rm Co}$  are the  lifetimes of  the corresponding  decays, and
$S^\gamma_{\rm Ni}$,  $S^\gamma_{\rm Co}$  and $S^\beta_{\rm  Co}$ are
the  respective   energies.   In  this  expression   the  $\gamma$-ray
deposition function can be well approximated by \citep{CPK80}:
\begin{equation}
f_{\rm dep}^\gamma=G\left(\tau\right)\left[1+2G\left(\tau\right)
\left(1-G(\tau)\right)\left(1-\frac{3}{4}G\left(\tau\right)\right)\right]
\end{equation}
and
\begin{equation}
G\left(\tau\right)=\frac{\tau}{\tau+1.6}
\end{equation}
being $\tau$ the optical depth. 

\begin{figure}[t]
\centering
    \includegraphics[width=0.6\columnwidth,angle=-90,clip]{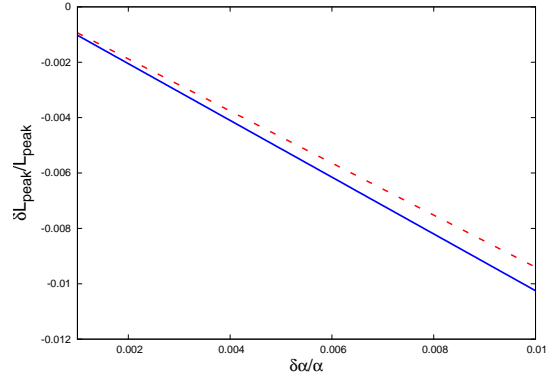}
    \caption{Peak  luminosity  of  distant  Type Ia  supernovae  as  a
      function of $\delta\alpha/\alpha$. The solid line corresponds to
      the  case  in which  both  the  variation of  the  Chandrasekhar
      limiting mass and the variation  of the opacity of the expanding
      photosphere are considered, while the dashed line corresponds to
      that in  which only the variation  of the opacity is  taken into
      account.}
\label{fig:dlda}
\end{figure}

We first  compute the time  at which  the peak luminosity  occurs.  At
this time the diffusion timescale, $t_{\rm diff}$ equals the expansion
timescale,  $t_{\rm  exp}$.   Hence,   we  have  $t_{\rm  peak}=t_{\rm
diff}=t_{\rm exp}$.   We now compute approximate  expressions for both
timescales.  The expansion timescale is  obtained from the velocity of
the ejected  material, $t_{\rm diff}=R/v$,  where the velocity  can be
obtained from the energy of the explosion:
\begin{equation}
v=\sqrt{\frac{2E}{M}}. 
\end{equation}
The diffusion timescale is given by:
\begin{equation}
t_{\rm diff}=\frac{\kappa\rho R^2}{c}
\end{equation}
where $\kappa\simeq$0.1~cm$^2$~g$^{-1}$ is the opacity. We substitute
the value of $\rho$ by its average value:
\begin{equation}
\rho=\frac{3M}{4\pi R^3}
\end{equation}
After some algebra we obtain:
\begin{equation}
t_{\rm diff}=\frac{3\kappa}{4\pi c v t_{\rm exp}}
\end{equation}
Taking into account that at $t_{\rm peak}$  the diffusion 
timescale and the expansion timescale are equal, we obtain
\begin{equation}
t_{\rm peak}=\left(\frac{3\kappa}{4\sqrt{2}\pi c}\right)^{1/2}
            \left(\frac{M^3}{E}\right)^{1/4}
\end{equation}
Here,   for  the   sake  of   simplicity,  we   will  only   focus  on
Chandrasekhar-mass  models.   Moreover,  we   will  assume  that  only
$\alpha$  varies,  and   that  the  values  of  $G$   and  $e$  remain
constant. Thus, both  $M$ and $E$ are determined  by the Chandrasekhar
limiting mass, and consequently depend  on $\alpha$. Also, the opacity
(mainly determined  by electron  scattering) depends  on the  value of
$\alpha$. Thus, we  have that a small variation of  the fine structure
constant, $\delta\alpha$, results in a  variation of the time at which
the peak luminosity occurs:
\begin{equation}
\frac{\delta t_{\rm peak}}{t_{\rm peak}}=\frac{1}{2}\frac{\delta\kappa}{\kappa}+
\frac{3}{4}\frac{\delta M}{M}-\frac{1}{4}\frac{\delta E}{E}
\end{equation}
Taking into  account the dependence  on $\alpha$  of $M$ and  $E$, and
assuming that  the opacity scales as  $\alpha^2$ \citep{ChibaKohri} we
finally obtain:
\begin{equation}
\frac{\delta t_{\rm peak}}{t_{\rm peak}}=-\frac{3}{8}\frac{\delta\alpha}{\alpha}
\label{deltat}
\end{equation}
We now investigate how $\tau$ scales on $\alpha$:
\begin{equation}
\tau=\kappa\rho R=\frac{3}{4\pi}\kappa\frac{M^2}{Et^2}
\end{equation}
At peak luminosity:
\begin{equation}
\tau=\sqrt{2}c\left(\frac{M}{E}\right)^{1/2}
\end{equation}
Consequently:
\begin{equation}
\frac{\delta\tau}{\tau}=\frac{1}{2}\frac{\delta M}{M}-
\frac{1}{2}\frac{\delta E}{E}=-\frac{7}{4}\frac{\delta\alpha}{\alpha}
\label{deltatau}
\end{equation}
Using this result we now study how $q$ depends on $\alpha$:
\begin{equation}
\frac{\delta q}{q}=-\frac{\delta t_{\rm peak}}{t_{\rm peak}}+
\frac{\delta f_{\rm dep}^\gamma}{f_{\rm dep}^\gamma}=
-\frac{\delta t_{\rm peak}}{t_{\rm peak}}+\eta\frac{\delta G}{G}
\label{deltaq}
\end{equation}
where
\begin{equation}
\eta=1+4G(t_{\rm peak})-10.5G(t_{\rm peak})^2+6G(t_{\rm peak})^3
\end{equation}
and
\begin{equation}
\frac{\delta G}{G}=1.6\frac{\delta\tau}{\tau}
\label{deltaG}
\end{equation}
Finally,      combining      Eqs.~(\ref{deltat}),      (\ref{deltaq}),
(\ref{deltatau}) and (\ref{deltaG}) we obtain the following expression
for the variation of $q$ at peak luminosity:
\begin{equation}
\frac{\delta L_{\rm peak}}{L_{\rm peak}}=
\frac{\delta q(t_{\rm peak})}{q(t_{\rm peak})}=
\left(\frac{3}{8}-\frac{7}{4}1.6\eta\right)
\frac{\delta\alpha}{\alpha}
\end{equation}
All in  all, it turns out  that the peak bolometric  magnitude, ${\cal
M}$, and hence  the luminosity distance of distant  SNIa are different
when a varying  $\alpha$ is considered.  The correction  to ${\cal M}$
is given by:
\begin{equation}
\delta {\cal M}=-2.5\frac{\delta L_{\rm peak}}{L_{\rm peak}}=
-2.5\left(\frac{3}{8}-\frac{7}{4}1.6\eta\right)
\frac{\delta\alpha}{\alpha}
\end{equation}
Note  that this  expression  differs from  that of  \cite{ChibaKohri},
because in addition to the term that accounts for the variation of the
opacity of the expanding photosphere there are terms which account for
the variation of  the mass of nickel synthesized  in the thermonuclear
outburst \citep{GG02}.  In Fig.~\ref{fig:dlda}  we compare our results
with those  of \cite{ChibaKohri}.   As can  be seen,  in our  case the
dependence  on $\delta\alpha/\alpha$  of  $\delta L_{\rm  peak}/L_{\rm
peak}$ is  steeper than that  of \cite{ChibaKohri}. This,  clearly, is
due to the fact that in our case  we do not only take into account the
dependence of the opacity on $\alpha$ and but also we consider that of
the mass of nickel synthesized  in the supernova outburst. However, we
stress that both  our results and those of  \cite{ChibaKohri} agree in
the fact that  a decrease of the value of  $\alpha$ translates into an
increase of the luminosity of thermonuclear supernovae. Thus a smaller
(larger) value of $\alpha$ makes SNIa brighter (fainter).

\subsection{The variation of $\alpha$}
\label{sec:dipole}

As mentioned, the data obtained using  the Keck and the VLT telescopes
during  the last  years has  resulted in  a set  of values  of $\Delta
\alpha/\alpha$ for $\sim 300$ absorption  systems covering most of the
sky.  This  extensive set  of data was  analyzed by  \cite{Webb11} and
\cite{King12}  and,  taken  together,  they concluded  that  there  is
evidence for an angular variation of $\alpha$. Moreover, they proposed
the following phe\-no\-me\-no\-lo\-gi\-cal model  for the variation of
$\alpha$:
\begin{equation}
\frac{\delta \alpha}{\alpha}=A+B  \cos\theta,
\label{alphavar}
\end{equation}
where $\cos\theta = \vec r \cdot \vec D$, $\vec D$ is the direction of
the dipole, $\vec r$ is the position  on the sky, $A$ is a constant (a
monopole  term) and  $B$ is  the amplitude  of the  dipole term.   The
values  of $A$,  $B$  and $\theta$  depend somewhat  on  the data  set
considered studied \citep{King12}.  We  nevertheless emphasize that in
all the  models of \cite{King12}  $\alpha$ depends on  right ascension
and declination, and  moreover that the direction of  the dipole seems
to  be  well  established,   pointing  towards  the  same  approximate
direction on  the sky.  Thus,  here, for  the sake of  conciseness, we
will only  analyze their best fit  model, for which the  amplitudes of
the  monopole   and  dipole   terms  are   respectively  $A=(-0.177\pm
0.085)\times  10^{-5}$  and $B=(0.97^{+0.22}_{-0.20})\times  10^{-5}$,
and  the  dipole  term   points  towards  right  ascension  $17.4^{\rm
h}\pm1.0^{\rm h}$ and declination $-61^\circ\pm10^\circ$.  In a second
step we  will consider  the effects  of a  varying $\alpha$  using the
results  obtained in  Sect.~\ref{sec:mbol}, but  leaving $A$,  $B$ and
$\vec D$ as free parameters, and we will obtain their values using the
observed data of Type Ia supernovae.

\subsection{Reference cosmological model}
\label{sec:lcdm}

We adopt  as a  reference model  to compare  with a  flat $\Lambda$CDM
model with  the following cosmological parameters.  The matter density
in units  of the critical  density is  $\Omega_{\rm M} =0.264$  and we
also  take  $\Omega_{\rm  R}=0$.   At last,  the  Hubble  constant  is
$H_0=71.2\,{\rm  Mpc^{-1}\, km  \, s^{-1}}$.   These are  the best-fit
values presented by the WMAP  collaborarion using the 9-year WMAP data
of the  Cosmic Microwave Background \citep{wmap9cls},  the temperature
power  spectrum for  high  $l$ from  the  Atacama Cosmology  Telescope
\citep{ACT13} and South Pole  Telescope \citep{SPT12}, the position of
the Baryon Acoustic Oscillations peak \citep{BAO1,BAO2,BAO3,BAO4}, and
the   three   year   sample    of   the   Supernovae   Legacy   Survey
\citep{Guy10,Conley11,Sullivan11}.

\begin{figure}[t]
\centering
    \includegraphics[width=0.6\columnwidth,angle=-90,clip]{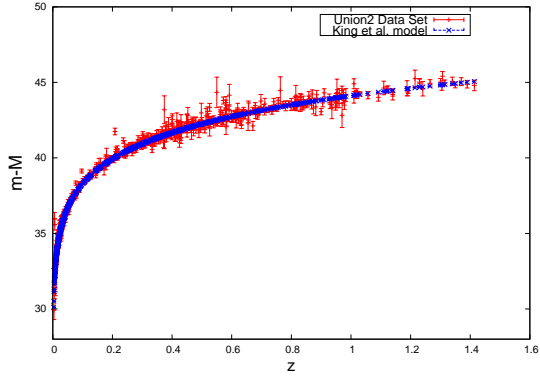}
    \caption{A comparison of the distance modulus and redshift for the
      model of  \cite{King12}.  The  observed data from  the Union~2.1
      compilation and their respective errors  are shown in red, while
      the theoretical  predictions are  shown using blue  symbols. See
      the online  edition of the journal  for a color version  of this
      figure.}
\label{modvsredshift}
\end{figure}

\section{Results}
\label{resultados}

\begin{figure*}[t]
\centering
    \includegraphics[width=0.25\textwidth,angle=-90,clip]{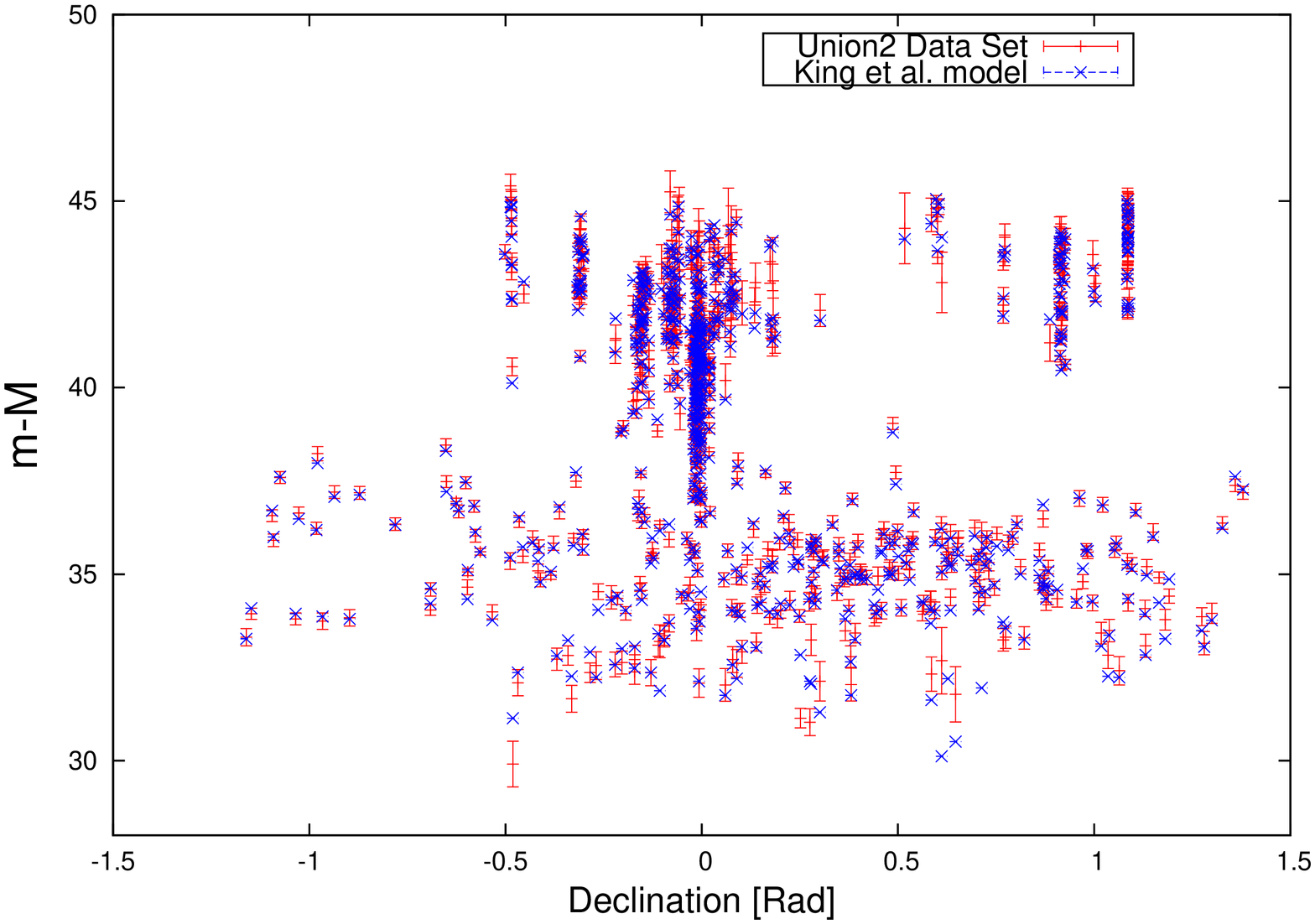}
    \includegraphics[width=0.25\textwidth,angle=-90,clip]{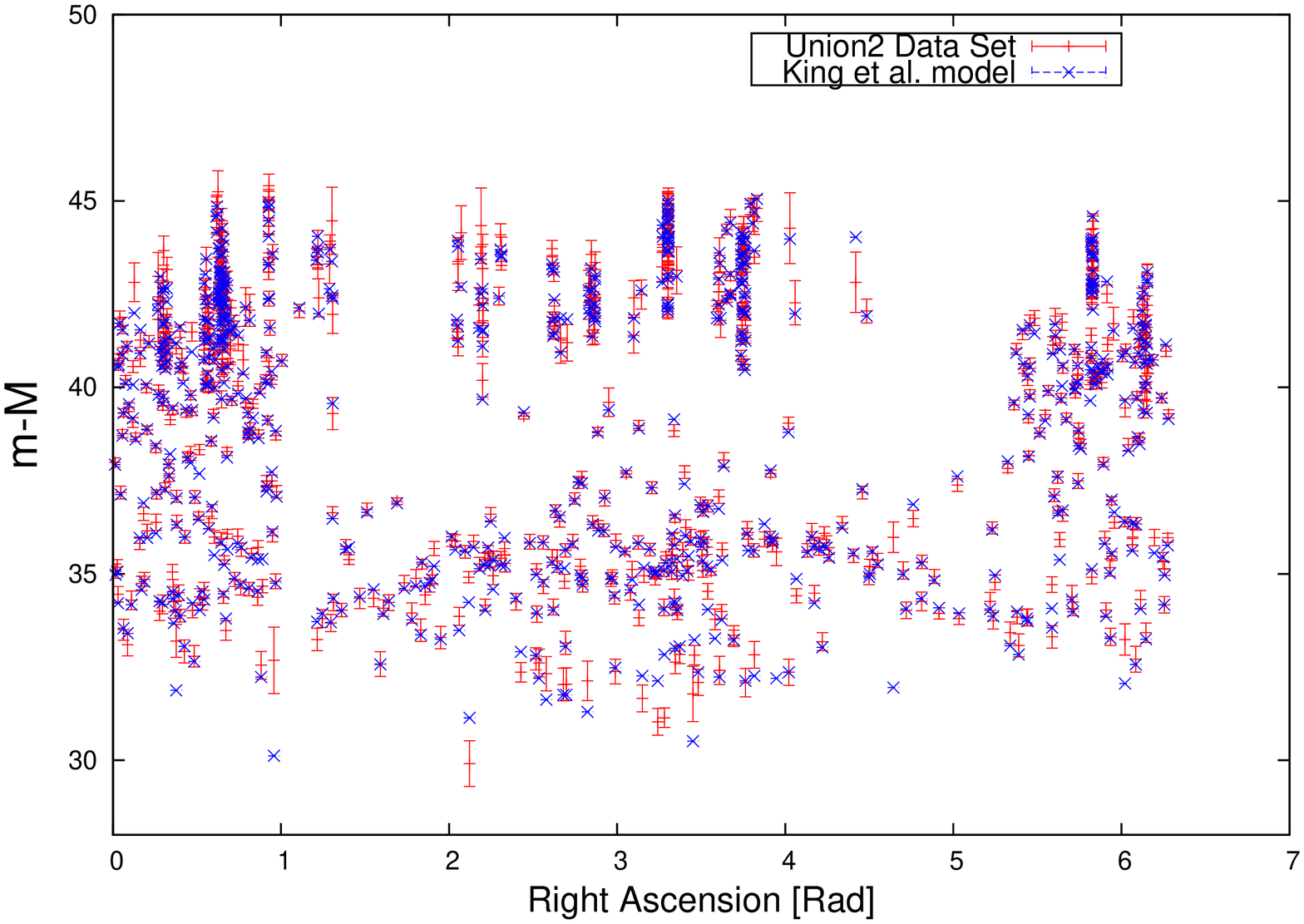}
    \caption{A comparison of the  distance modulus and right ascension
      (left panel)  and declination  (right panel),  for the  model of
      \cite{King12}.   Again, the  observed data  and the  theoretical
      ones are shown using red and blue symbols.}
\label{modvsardec}
\end{figure*}

In  this section  we show  the results  of comparing  the data  of the
Union~2.1 compilation  of SNe  Ia with  the phenomenological  model of
\cite{King12}.  To  this end,  in Fig.~\ref{modvsredshift}  we compare
the relation between\-  the distance modulus and the  redshift for the
theo\-re\-ti\-cal model  and the  observational data of  the Union~2.1
compilation.  As can be seen,  the theoretical model matches very well
the  observated  luminosity  distance-redshift relationship.   We  now
check whether there is an angular  dependence of the value of the fine
structure constant.  Fig.~\ref{modvsardec}  shows the distance modulus
as  a function  of the  right ascension  (left panel)  and declination
(right  panel)   of  the   absorption  systems,   for  the   model  of
\cite{King12} considered here (blue points) and the observational data
(red points).  Again, overall all  the phenomenological model seems to
explain well most of the  observed supernovae, although there are some
differences  for each  individual  SNIa, depending  on its  respective
position  in the  sky.  Moreover,  it  can be  seen that  there is  no
obvious correlation between the value  of the distance modulus and the
position in the sky.

Since the amount of available observational data is sufficently large,
it is crucial to further quantify  the degree of agreement between the
observed data and the theoretical models.   To do so we use a $\chi^2$
test.   The  $\chi^2$ estimator  is  constructed  using the  following
expression:
\begin{equation}
\chi^2 = \sum \frac{\left[(m(z,\theta) - M_0)_{\rm P} - 
(m - M_0)_{\rm R}\right]^2}{\sigma_{\rm O}^2}
\end{equation}
In this equation $(m(z,\theta) - M_0)_{\rm P}$ is computed considering
the   hypothetical   variation   of    $\alpha$   according   to   the
phenomenological model of \cite{King12} and considering the results of
Sect.~\ref{sec:mbol},  whereas $(m  - M_0)_{\rm  R}$ and $\sigma_{\rm O}$ 
are the observational data and the observational errors of the distance modulus
taken both from the  Union~2.1 compilation.   We obtain  that the  reduced
$\chi^2$ --  that is the  value of $\chi^2$  divided by the  number of
degrees of freedom,  $\nu$ -- for the  phenomenological model proposed
by \citet{King12} is $\chi^2/\nu=1.74591$, while for the case in which
no variation of $\alpha$ is considered we obtain $\chi^2/\nu=1.74589$.
Thus, the differences  are not statiscally significant.   We note that
when the  complete data set  of the Union~2.1 compilation  ($713$ data
points) is used, the reduced value of $\chi^2$ is slightly larger than
expected in  both cases.  This  is due to  the fact that  although the
vast majority of the data fit very well with our standard model, there
are  some supernovae  that  do  not.  This  issue  has been  discussed
previously in the literature \citep{ala}, and thus we will not discuss
it in detail here.  Instead, we refer to the previously mentioned work
for an extensive discussion of the  problem, and we simply discard the
17 conflictive  data points that  \cite{ala} recommend to do  not use.
When  this procedure  is  adopted we  obtain $\chi^2/\nu=1.03494$  and
$\chi^2/\nu=1.03493$, respectively.

\begin{table*}[t]
\begin{center}
\begin{tabular}{cccccc}
    \hline
    \hline
     Model & $A$ & $B$ & R.A. (hr) & $\delta$ ($^{\circ}$) & $\chi^2/\nu$ \\  
     \hline
     1  & $(1.141\pm0.297)\times10^{-2}$ & $(2.182\pm0.718)\times10^{-2}$ & $23.013\pm2.052$ & $(65.911\pm10.512)$ & 1.681\\
     2  & $(7.811\pm2.821)\times10^{-3}$ & $(2.122\pm0.785)\times10^{-2}$ & $1.313\pm4.268$ & $(75.719\pm10.052)$ & 1.001\\
    \hline
   \end{tabular}
\caption{Parameters of  the dipole  for the different  models obtained
  from the statistical analysis.}
  \end{center}
\label{tab:inverse}
\end{table*}

It is  nevertheless interesting to  go one  step beyond and  adopt the
inverse procedure. That is, check whether  or not there is a preferred
direction in the  raw observational data.  Hence, in a  second step we
consider $A$,  $B$ and  $\vec D$  as free  parameters, and  obtain the
resulting  values  using  the  observational  data  of  the  Union~2.1
compilation,  this time  employing  the  luminosity distance  computed
according to the results of Sect.~\ref{sec:mbol}.  We will do so using
the complete Union~2.1 data set (model 1) and the reduced data set, in
which only $696$ data points are considered (model 2).  The results of
this  exercise are  shown in  Table 1. As  it happened
when considering the models of  \citet{King12}, the values of $\chi^2$
for model 1  is larger than expected while we  find a reasonable value
when  only  $696$   data  points  are  included   in  the  statistical
analysis. Moreover,  it follows from Table 1 that the
values of $A$, $B$ and $\vec D$ are considerably different for the two
sets  of data  studied  here.   In particular,  the  amplitude of  the
monopole  term  ($A$)  is   significantly  larger  when  the  complete
Union~2.1  dataset is  employed,  and moreover  for  both datasets  we
obtain  values that  are  considerably larger  than  that obtained  by
\cite{King12}  employing  the  many  multiplet  method,  and  that  of
\citet{YWC14} using  Type Ia supernovae, but  disregarding the effects
of  a varying  $\alpha$.   However,  we remark  that  given the  large
uncertainties in the  determination of $A$ our  results are compatible
with  a null  result for  model 2,  which is  obtained using  the more
reliable data.  Also the direction of  the dipole term is different in
both  cases,   although  their  respective  amplitudes   are  similar.
Moreover, the direction  of the dipole when the  correct dependence on
$\alpha$  is   considered  is   at  variance   with  the   results  of
\cite{King12} for distant quasars and \citet{YWC14} for SNIa.

\section{Discussion and conclusion}
\label{discusion}

In this paper we have studied whether the recently reported space-time
variation  of  the  fine  structure  constant  \citep{King12}  can  be
confirmed   or   discarded   using  the   Union~2.1   compilation   of
lu\-mi\-no\-si\-ty distances of  SNIa.  To do so we  have derived from
simple  physical  arguments a  scaling  law  for the  peak  bolometric
magnitude  of  distant SNIa.   Our  results  show that  the  currently
a\-vai\-la\-ble data does not allow  to either confirm nor discard the
phenomenological  models  of  \cite{King12}  and  \cite{Webb11}.   The
ultimate  reason for  this  is  that the  magnitudes  of the  reported
variations  of  $\alpha$  result  in modest  variations  of  the  peak
bolometric magnitudes of distant SNIa, and thus the differences in the
positions of the SNIa of the  Union~2.1 compilation are too small when
compared with the leading terms  intervening in the calculation of the
lu\-mi\-no\-si\-ty distances of Type Ia  supernova. To this regard, it
is worth mentioning  that \citet{YWC14} have found that  the SNIa data
can  be  better  explained  when   a  dipole  model  pointing  towards
$(b=-14.3^\circ\pm10.1^\circ,\,   l=307.1^\circ\pm16.2^\circ)$  --   a
direction close  to that  found by  \cite{King12}.  However,  in their
calculations they did not include  the effects of a possible variation
of $\alpha$,  and instead assumed  that all the  fundamental constants
were indeed truly  constant. Our approach goes one step  beyond and we
included it.   In a second step  we used the Union~2.1  compilation to
check whether or not there exists a variation of $\alpha$, and we have
found that the monopole term  cannot be determined with accuracy given
the  still large  uncertainties,  and  that for  the  dipole term  the
direction is at  odds with that found in previous  studies.  Thus, the
analysis performed  here shows that  if such a preferred  direction in
the  SNIa data  of the  Union~2.1  catalog exists,  its origin  cannot
likely be  due to an eventual  variation of $\alpha$.  In  summary, we
conclude  that the  actually available  SNIa  data cannot  be used  to
distinguish between a standard cosmological model in which $\alpha$ is
strictly  constant  and  a  model  where  $\alpha$  has  a  space-time
variation.


\acknowledgements Support  for this work  was provided by  PIP 0152/10
grant, by CONICET, by MCINN  grant AYA2011--23102, and by the European
Union FEDER funds.



\end{document}